# Digital Cardan Grille: A Modern Approach for Information Hiding

Jia Liu[1], Tanping Zhou[1], Zhuo Zhang[1,2], Yan Ke[1], Yu Lei[1]
Minqing Zhang[1], Xiaoyuan Yang[1]

[1]Key Laboratory of Network and Information Security of PAP, Engineering University of PAP, Xi'an, 710086, China
[2] Rocket Force University of Engineering, Xi'an 710000,China
`liujia1022@gmail.com, 850301775@qq.com`

**Abstract.** In this paper, a new framework for construction of Cardan grille for information hiding is proposed. Based on the semantic image inpainting technique, the stego image are driven by secret messages directly. A mask called Digital Cardan Grille (DCG) for determining the hidden location is introduced to hide the message. The message is written to the corrupted region that needs to be filled in the corrupted image in advance. Then the corrupted image with secret message is fed into a Generative Adversarial Network (GAN) for semantic completion. The adversarial game not only reconstruct the corrupted image , but also generate a stego image which contains the logic rationality of image content. The experimental results verify the feasibility of the proposed method.

**Keywords:** Information Hiding, Cardan grille, Steganography , Generative Adversarial Network

## 1  Introduction

In 1550, Girolamo Cardano (1501-1576), known in French as Jérôme Cardan, proposed a simple grid for writing hidden messages. He intended to cloak his messages inside an ordinary letter so that the whole would not appear to be a cipher at all. Such a disguised message is considered to be an example of steganography, which is a sub-branch of general cryptography. For a long time, the difficulty of constructing the 'ordinary letter' made 'steganography' nearly going to be 'information modifying'. These traditional steganography methods such as [1] [2], by modifying the cover data to hide the information, had to struggle against the steganalysis technique[3].

In this paper, we come back to the road of Cardan, a new generative image steganography framework is proposed based on semantic inpainting. Firstly, the corrupted image is taken as the cover, and the secret information is written to the area that needs to be filled by a digtal Cardan grille. In the case of keeping the secret message unchanged, the semantic image completion is realized by using the generative adversarial network. The secret message is hidden in the reconstructed image after completion. The experiments on the image database confirms the validity of such simple method.

## 2      Related Work

Fridrich [4] etc. discuss the cover selection and synthesis method for information hiding. The concealer selects the image in the normal image base according to the secret information, and the receiver calculates the image Hash to get the secret information. This is a non modified information hiding technology, which does not modify the cover image, thereby avoiding the threat of the existing steganalysis technology. This method can not be applied to practical applications because of its low payload. With the help of texture synthesis, [5,6] are proposed for data hiding. They use the sample texture and a bunch of color points generated by secret messages to construct dense texture images. [7] improves the embedding capacity by proportional to the size of the stego texture image. Qian etc. [8] proposes a robust steganography based on texture synthesis. Similar to these method，Xue etc. [9] use marbling, a unique texture synthesis method that allows users to deliver personalized messages with beautiful, decorative textures for hiding message. This kind of texture-based steganography is based on the premise that the cover may not represent the content in real world . Similar to [3], Zhou etc.[10] proposed a method called Coverless Information Hiding based on the bagof-words model (BOW) . According to the mapping relation, a set of sub-images with visual words related to the text information is found. The images containing these sub-images are used as stego images for secret communication. [11] surveys these steganographic techniques using the two categories of data as the camouflage, including semi-creative steganography and creative steganography.

Generative adversarial networks (GANs) have become a new research hot spot in artificial intelligence. The aim of GAN is to estimate the potential distribution of existing data and generate new data samples from the same distribution. Since its initiation, GAN has been widely studied due to its enormous prospect for applications, including image and vision computing, speech and language processing, information security. Recently, Two types of designs have applied adversarial training to cryptographic and steganographic problems. Abadi [12] used adversarial training to teach two neural networks to encrypt a short message, that fools a discriminator. However, it is hard to offer an evaluation to show that the encryption scheme is computationally difficult to break. Adversarial training has also been applied to steganography. Volkhonskiy etc.[13] first propose a new model for generating image-like containers based on Deep Convolutional Generative Adversarial Networks (DCGAN[14]). This approach allows to generate more setganalysis-secure message embedding using standard steganography algorithms. Similar to [13]，Shi  etc.[15] introduce a new generative adversarial networks to improve convergence speed, the training stability the image quality. Similar to the Abadi [12], [16] define a game between three parties, Alice, Bob and Eve, in order to simultaneously train both a steganographic algorithm and a steganalyzer. Tang etc.[17] propose an automatic steganographic distortion learning framework using a generative adversarial network, which is composed of a steganographic generative subnetwork and a steganalytic discriminative subnetwork. However, most of these GAN-based steganographic schemes are still the steganographic techniques for modifying images.

Since GAN's biggest advantage is to generate samples, it seems that it is a very intuitive idea to use GANs to generate a semantic stego carrier from a message directly. However, the extraction of message is an important constraint to the information hiding by GANs. Some researcher have made a preliminary attempt on this intuitive idea. Ke [18] proposed generative steganography method called GSK in which a the secret messages are generated by a cover image using a generator rather than embedded into the cover, thus resulting in no modifications in the cover. Liu etc.[19] propose a method that use ACGANs [20] to classify the generated samples, and they make the class output information as the secret message. We call these methods generative steganography.

As far as we know, we are the first try to bulid a digtal cardan grille method for generative steganography. This paper has the following contributions:

1. Compared with texture-based methods [8] and image-text mapping methods[10], semantic image completion technology is used to ensure the logical rationality of cover contents while correctly representing secret information.

2. This method can be regarded as the automated version of the Carden-grille method, since the traditional Carden-grille method needs to construct the seemingly meaningful cover based on the secret message, such as the *Tibetan poem*, which takes time and effort. The method of this paper relies on secret messages, which are automatically filled (hidden) by generative model. It can also be extended to other media, such as text, video and other fields.

3. This method propose a framework of generative steganography with Kerckhoffs' principle. All processes can be made public except for the parameter of the Carden-grille (as secret keys that are shared by both parties). Ideally, without shared keys, the extraction of secret information is equivalent to brute force cracking.

## 3      Digital Cardan Grille for Imformation Hiding

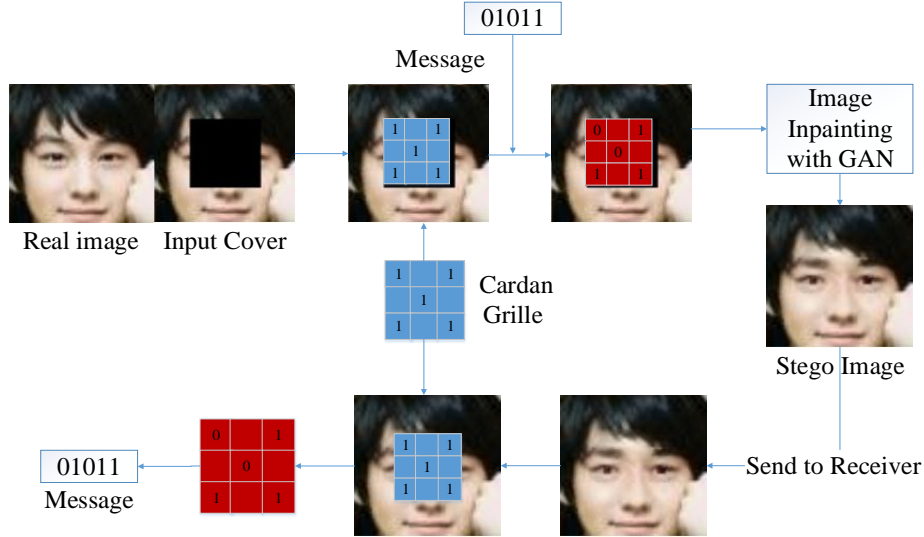

**Fig** 1. The proposed Framework for infomaion hiding using Cardan grille.

In our framework , as illustrated in Fig.1, the process of information hiding is in line with the basic idea of traditional Cardan grille. The sender defines a mask, called Digtal Cardan grille, to determine where the message is hidden, and the secret messages go directly to these locations of the input cover image which is corrupted. Then, an image inpainting method based on GANs is used to finish the image completion. A well-filled image is transmitted to the recipient through the public channel. The receiver extracts a secret message using the Cardan grille shared by the two parties in the reconstructed image. The core of this framework is to define Cardan grille method that not only ensure the consistency of the secret messages but also the logical rationality of cover contents using GANs before and after the completion of the image inpainting. The development of the GANs has given the technical foundation to satisfy these two premises. In the next section, we will give the details of information hiding, image inpainting, and message extraction. In fact, this framework describes a general automated Cardan grille, and the input cover vector can be a piece of text, image, video, and other type of media.

### 3.1      Information Hiding with Cardan Grille

Information hiding method based on Cardan grille, can be seen as a processing to build a mapping relationship between the message *m* and stego *s*, Cardan grille is used for message extraction. In this section, we first try to formalize the Cardan grille method. Then, we give a practical approach for information hiding.

The principle of Cardan grille algorithm is to create or generate a new stego carrier $s_{fake}$ which is not exist in the real word from $m$ for any given Cardan grille. The fake stego carrier should satisfy these constrains as follow:

$$s_{fake} = G(m) \tag{1}$$

$$m = C(s_{fake}) \tag{2}$$

$$p_{fake} = p_{real} \tag{3}$$

where $G(.)$ is a generator, $s_{fake}$ denotes the fake stego carrier, $C(.)$ is the extraction operation using Cardan grille, $p_{fake}$ and $p_{real}$ denotes the distribution of fake stego carrier and real cover. (1) and (2) make the Cardan grille method close to cardboard ciphers algorithm. Ideally, condition (3) can be realized by a powerful generative model such as GANs which make the Cardan grille became a steganography problem.

Unfortunately, design a generator satisfying all these constrains is still a difficult problem. In this paper, the process of generating the stego is decomposed into two steps to simplify the designing. First of all, we can define a mapping operation:

$$m' = E_{CG}(m) \tag{4}$$

that makes the secret message $m$ by Cardan grille which is shared by both parties to a new expanded message $m'$ subject to follow constrain:

$$m = C(m') \tag{5}$$

Then, we can get the stego carrier by:

$$s_{fake} = G(m') \tag{6}$$

We will show that this simple separation trick make it easier to build a generator.

With the help of the thought mentioned above, we expand the message $m$ to $m'$, shown in the fig2 below.

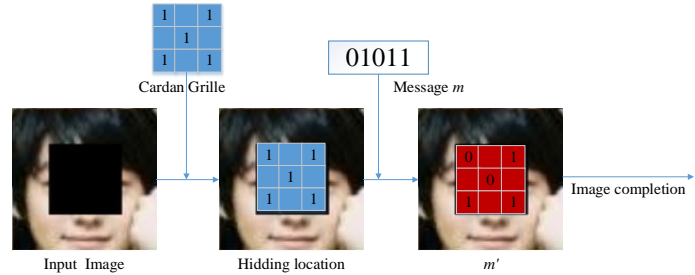

**Fig 2.** Flowchart of Message preprocessing.

First, select the secret input corrupted image $y_{corrupted}$, message $m$, and Cardan grille. It's important to note the size and the value of this Cardan grille, the location in the corrupted image are shared by both parties. Assume that the size of the corrupt region is $a*b$, where $a=b=3$ ($a=b=32$ in fact) in Fig2. Then a Cardan grille with same size is defined as:

$$M_{CG} = \begin{bmatrix} 1 & 0 & 1 \\ 0 & 1 & 0 \\ 1 & 0 & 1 \end{bmatrix} \quad (7)$$

We use a binary mask that has values 0 or 1. A value of 1 represents the parts of the region we want to hide message and a value of 0 represents the parts of the image we can not write message. This binary mask is similar to the wet paper code which is a tool for constructing steganographic schemes. The location of the Cardan grille is at the center of the damaged image. Then the message '01011' can be written in the input image. We get a corrupted image contains secrete message shown as $m'$. Note that $m = m' \odot M'_{CG}$, where $M'_{CG}$ is zero padding version of $M_{CG}$, $\odot$ denotes the element-wise product operation.

In fact, as shown in Fig.2 , we only solved the problem of message extension, but this extension is so important that it transforms the image completion into information hiding. In the next section, we will give the details for the image completion based on the GANs, which together complete the entire information hiding process.

### 3.2 Semantic Inpainting Image for Information Hiding

As mentioned above, the image completion used for information hiding should satisfy two objectives, one is the rationality of the complete image content, the other is the stability of the message. In this paper we use the a image inpainting method which proposed by Yeh [21] based on a Deep Convolutional Generative Adversarial Network (DCGAN).

We use a binary mask $M$ that has values 0 or 1. A value of 1 represents the parts of the image we want to keep and a value of 0 represents the parts of the image we want to complete. Suppose we've found an image from the generator for some that gives a reasonable reconstruction of the missing portions. The completed pixels can be added to the original pixels to create the reconstructed image:

$$x_{reconstructed} = M \odot y + (1 - M) \odot G(\hat{z}) \quad (8)$$

The contextual and perceptual information in [21] are used for defining loss functions.

**Contextual Loss**: To keep the same context as the input image, make sure the known pixel locations in the input image $y$ are similar to the pixels in $G(z)$. We need to penalize $G(z)$ for not creating a similar image for the pixels that we know about. Formally, we do this by element-wise subtracting the pixels in $y$ from $G(z)$ and looking at how much they differ:

$$L_{contextual}(z) = || M \odot G(z) - M \odot y ||_1 \quad (9)$$

In the ideal case, all of the pixels at known locations are the same between $y$ and $G(z)$. Then $G(z)_i - y_i = 0$ for the known pixels $i$ and thus $L_{contextual}(z) = 0$.

**Perceptual Loss**: To recover an image that looks real, let's make sure the discriminator is properly convinced that the image looks real. We'll do this with the same criterion used in training the DCGAN:

$$L_{perceptual}(z) = \log(1 - D(G(z))) \quad (10)$$

Contextual Loss and Perceptual Loss successfully predict semantic information in the missing region and achieve pixel-level photorealism.

**Message Loss**: The key point of using image completion for information hiding is that the messages generated by the mask $M_{CG}$ in Cardan Grille should be as stable as possible. The pixel value of the corresponding position of the generated image is equal to the value of the secret message.

$$L_{message}(z) = || M'_{CG} \odot G(z) - M'_{CG} \odot m' ||_1 \tag{11}$$

In the ideal case, all of the pixels at hiding locations are the same between $m'$ and $G(z)$. Then $G(z)_i - m'_i = 0$ for the known pixels $i$ and thus $L_{message}(z) = 0$. In practice, for each 8 bits pixel point on each layer, we only use the first several bits, as messages, and the some lower significant bits as redundancy that can be used to ensuring the stability of embedded messages. Here we define a stability index *SI*, *SI* =5 means the 5 lower significant bits as redundancy.

We're finally ready to find $\hat{z}$ with a combination of the all these losses:

$$L(z) = L_{contextual}(z) + \lambda L_{perceptual}(z) + L_{message}(z) \tag{12}$$

$$\hat{z} = \arg\min L(z) \tag{13}$$

where $\lambda$ is a hyper-parameter that controls how import the contextual loss is relative to the perceptual loss.

### 3.3 Message Extraction

Message extraction for the receiver's is simple, the basic process is as shown in the Fig.3 below:

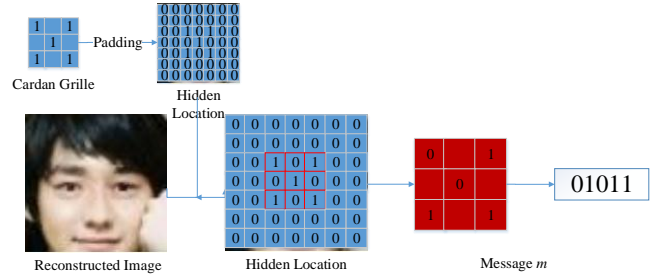

**Fig 3.** Message Extaction using Cardan grille

The receiver will cover the Mask directly on the image after reconstruction, and the secret message of the corresponding position can be obtained. The basic operation is as follows:

$$m = x_{reconstructed} \odot M'_{CG} \tag{14}$$

where $M'_{CG}$ is obtained by zero padding to $M_{CG}$.

## 4    Experiments

As a proof of concept, we implemented our adversarial training scheme on the LFW datasets [22]: a database of face photographs designed for studying the problem of unconstrained face recognition. The data set contains more than 13,000 images of faces collected from the web. We use alignment tool to pre-process the images to be 64x64,as shown in Fig.4 . We used the DCGAN model architecture from Yeh et al. [21] in this work. I emphasize that we modifies Brandon Amos's bamos/dcgan-completion.tensorflow [23] for information hiding. 12000 samples are used for training DCGAN. Our setting for image completion is same as the Brandon Amos's. The location of Cardan grille, added one for the center portions of corrupted images. The size of grille is fixed as 32*32 which is same as the size of corrupted region. We intentionally randomize the secret message so that the stability of embedded messages can be given in a quantitative manner.

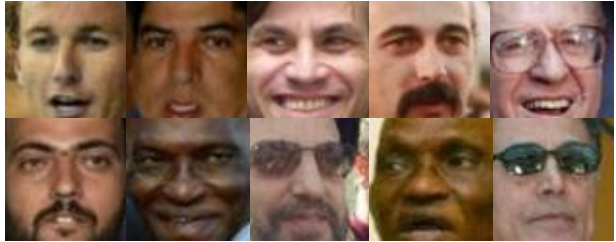

**Fig**. 4. Aligned samples form LFW database

Our results are shown in Fig. 5, which demonstrate that our method can successfully predict the missing content. It's important to emphasize that, in our experiment, gardan grille was randomly generated, and, in all the places that we could hide for every level in the image, we wrote the message which is also randomly generated.

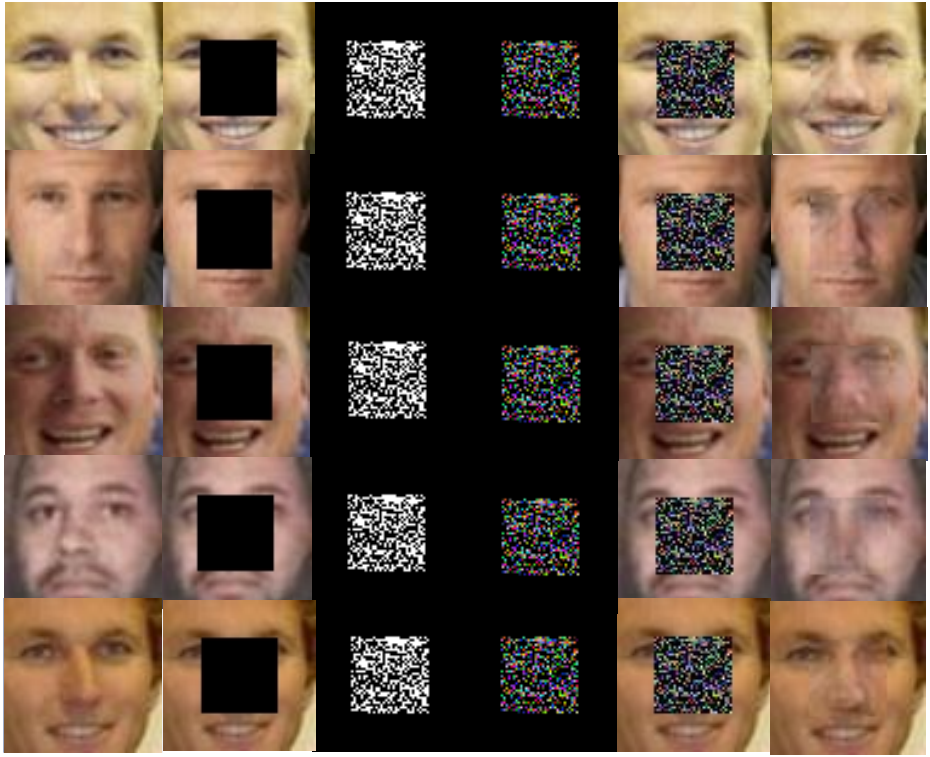

**Fig. 5.** For each example, **Column** 1: Original images from the dataset. **Column** 2: Images with central region missing. **Column** 3: Cardan grille with zero padding . **Column** 4: Secrete message with zero padding. **Column** 5: Corrupted Image with message. **Column** 6: Inpainting of column 5 by our method.

We also show the completion image generation process in Fig.6 with the number of iterations from 60 to 600. It can be seen that the complemented image becomes more real as the number of iterations increases.

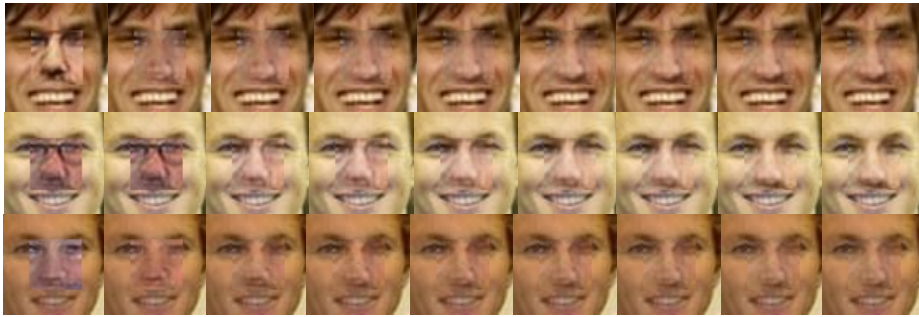

**Fig. 6.** The process of image generation.

Fig .7 shows the effects of different sizes of Cardan grille on the generation of images. From top to bottom, the grid sizes are 8, 16, 32, 48, respectively. As can be seen ,with the size increases, the generated image does have some distortion, especially at the edge of the lattice. For the case that the cardan grille size exceeds the completion mask (shown in row 4 in Fig 7. ), The image has more noticeable distortion

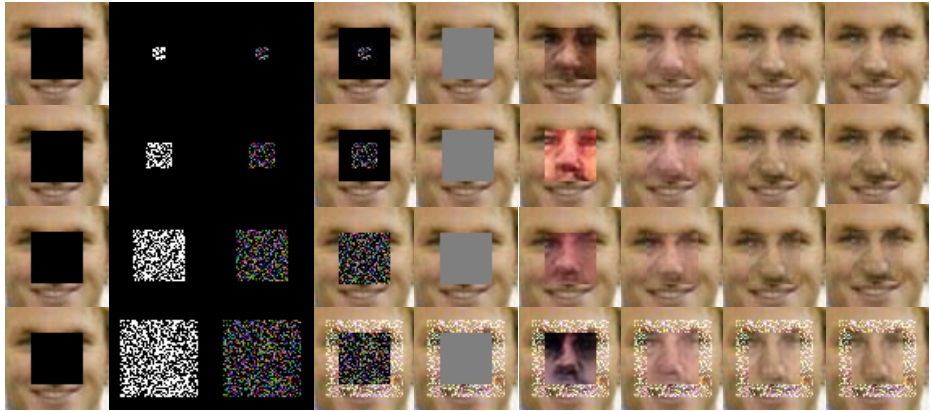

**Fig. 7.** Information hiding with diferent sizes of Cardan grille. Completion mask is shown in **Column** 5.

We also showed that the corrupted image is directly fed into the image completion network (see Fig. 8), which means that all embedded messages are 0. As the result can be seen, as the number of iterations increases, in order to satisfy the message loss constraint,the center of the completion image is closer and closer to the input image, Note that the contents of the filled parts of the completion region (appearance of the nose) can be identified.

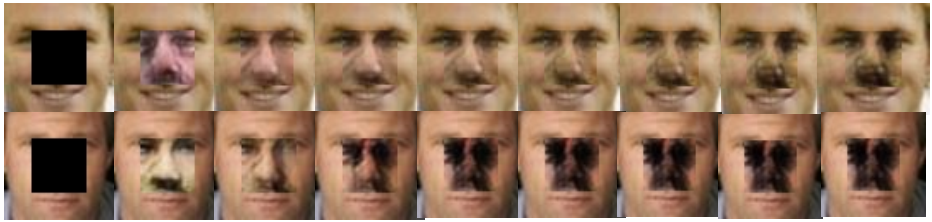

**Fig**. 8  Results on corrupted image.

Next, we show the results of information extraction, and Fig. 9 shows the whole process of information hiding. Notice that the last two images show extracted message and the error between the extracted message and the hidden message.

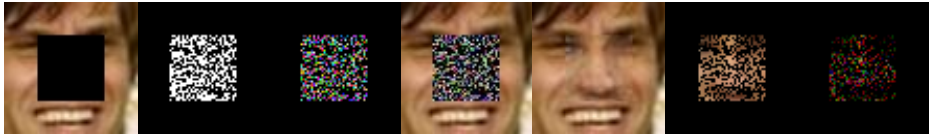

**Fig**. 9 Information extraction

**Fig**. 10 shows the relationship between the error rate of the message extraction and the number of iterations with different stability indices (4-7). With the increase of the stability index, the error rate of message extraction is reduced, but we need to point out that the best error rate is 0.42 when the stability index is 7. This is mainly due to the fact that in the previous experiment, gardan grille was randomly generated, about 50% of the area used for message hiding,all the pixel used for hiding is generated by GANs.

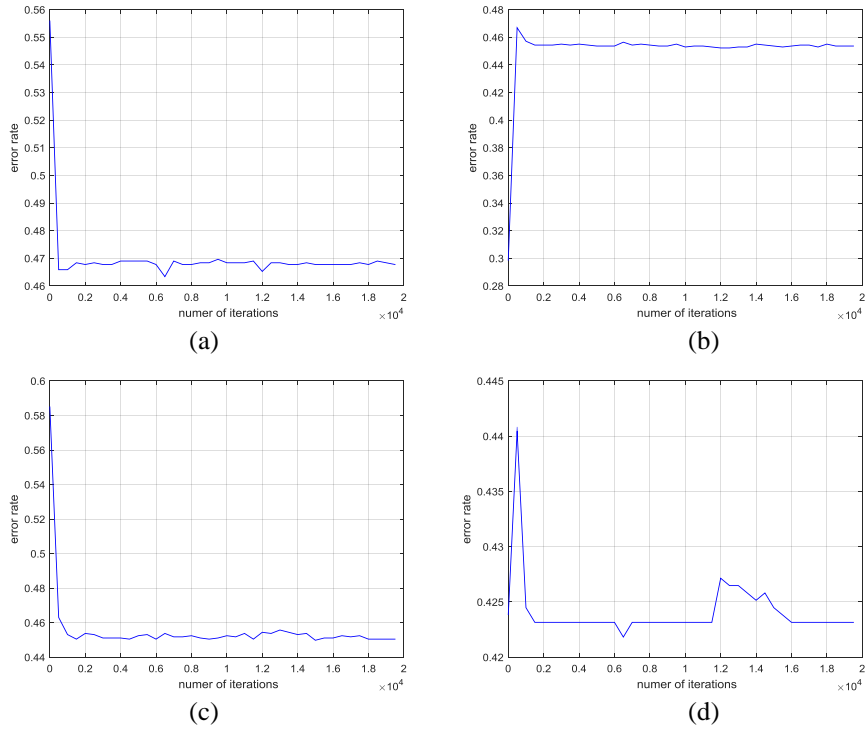

**Fig.** 10 Error rate for stability index with 4,5,6 and 7 respectively.

For the position that we can write on grille, we set the completion mask $M = 1$, which is not to be filled in this position. This keeps the message stable, and Fig.11 shows the results of generated images with 2000 iterations. It should be noted that the fourth subgraph of **Fig**.11 (a) shows that the hidden image have heavy distortion. This is mainly due to the fact that we write message at about 50% of the corrupted area. **Fig** 11.(b) shows the result of writing a small amount of information in the image.

The distortion of the hidden image is very small, and the message can be extracted correctly.

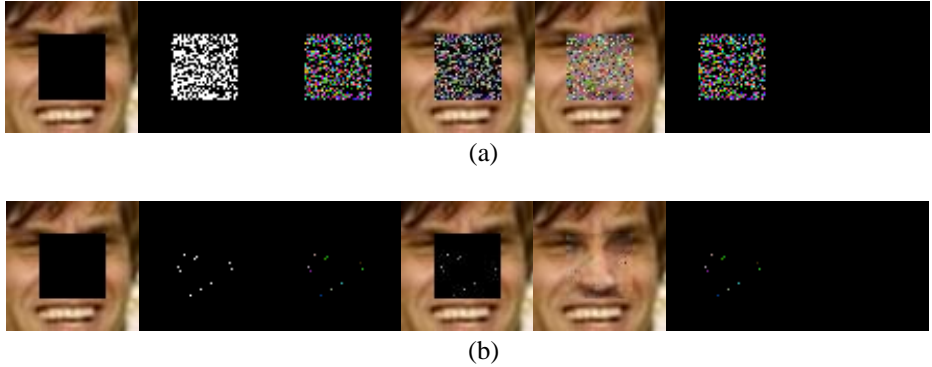

(a)

(b)

**Fig**. 11  All the pixels on the grille used for hiding are not to complete.

## 5    Conclusion and Future Work

In this paper, a method to build the Carden grille framework is proposed, which uses the semantic image completion technique to generate the stego image directly driven by secret message. Compared with the texture-based method and image-text mapping method, the logic rationality of the content of the stego image is guaranteed. This scheme can be extended to other media, such as text, video and other fields. It can also be seen as a method of classical cryptography, the safety depends on the parameters of the Cardan grille, without it, the extraction of the secret information in essence equivalent to brute force. This paper provides a basic framework for information hiding research based on generation model, and the experimental results verify the of promising  this method.

It is important to note that in this article, we assume that the message loss and Contextual loss is equally important, when two completely unrelated, image generation is better, when both completely relevant, information extraction accuracy is high, but the content of the image has obvious distortion. In our future work, we would like to pay more attention to the relationship between the message loss and contextual loss, looking for a better balance between the two losses.

### Acknowledgments
This work was supported by National Key R&D Program of China (Grant No. 2017YFB0802000), National Natural Science Foundation of China (Grant Nos. U1636114, 61772550, 61572521, 61379152 and 61403417), National Cryptography Development Fund of China (Grant No. MMJJ20170112）.